%% file: arXiv.tex
\documentclass[a4paper,11pt]{article}

\usepackage{hyperref}%
\hypersetup{%
  colorlinks=true,%
  linkcolor=blue,%
  citecolor=magenta,%
  urlcolor=black,%
  bookmarksnumbered=true,%
  bookmarkstype=toc,%
  bookmarksopen=false,%
  pdftitle={Discrete Scale Invariance and U(2) Family of Two-Body Contact Interactions in One Dimension},%
  pdfkeywords={discrete scale invariance; two-body problem; nonidentical particles},%
  pdfauthor={Satoshi Ohya}}%

\usepackage[margin=1in]{geometry}%

\usepackage[tt=false,sb]{libertine}%
\usepackage[T1]{fontenc}%
\usepackage{libertinust1math}%
\usepackage[cal=stix,scr=boondoxo,bb=boondox]{mathalpha}%
\usepackage{bm}%
\usepackage{mathtools}%
\DeclareMathOperator{\e}{e}%
\DeclareMathOperator{\supp}{supp}%
\allowdisplaybreaks[1]%

\usepackage{epic,eepic,xcolor}%

\usepackage[font=small,hang,labelfont=bf]{caption}%

\usepackage[title]{appendix}

\usepackage[backend=biber,style=phys,biblabel=brackets,block=ragged,eprint=true,url=true]{biblatex}%
\title{\Large\bfseries\boldmath Discrete Scale Invariance and $U(2)$ Family of Two-Body\\ Contact Interactions in One Dimension}%
\author{\normalsize Satoshi Ohya\\[1em]
  \small\itshape Institute of Quantum Science, Nihon University,\\
  \small\itshape Kanda-Surugadai 1-8-14, Chiyoda, Tokyo 101-8308, Japan\\[1ex]
  \small\ttfamily ohya.satoshi@nihon-u.ac.jp}%
\date{\small(Dated: \today)}%

\begin{document}
\maketitle%
\flushbottom%

\begin{abstract}
  Because of the absence of indistinguishability constraint,
  interparticle interactions between nonidentical particles have in
  general much more variety than those between identical particles. In
  particular, it is known that there exists a $U(2)$ family of
  two-body contact interactions between nonidentical particles in one
  spatial dimension. This paper studies breakdown of continuous scale
  invariance to discrete scale invariance under this $U(2)$ family of
  two-body contact interactions in two-body problems of nonidentical
  particles on the half line. We show that, in contrast to the
  corresponding identical-particle problem, there exist two distinct
  channels that admit geometric sequences of two-body bound states.
\end{abstract}

\section{Introduction}
\label{section:1}
Discrete scale invariance is an invariance under enlargements and
reductions of the system size by constant factors that form a
geometric sequence. If a quantum many-body system enjoys this
invariance, bound-state energies of many-body bound states, if exist,
must form a geometric sequence. The most prominent example that
realizes this phenomenon is the three-body problem of identical bosons
in three dimensions under two-body short-range interactions. This
example, first discovered by Efimov in 1970 \cite{Efimov:1970zz}, has
now been experimentally observed and also theoretically generalized to
several directions; see, e.g., \cite{Naidon:2016dpf} for a recent
review. The purpose of this paper is to present a new discrete
scale-invariant model of nonidentical particles in one dimension. The
key is the absence of indistinguishability constraint in
nonidentical-particle problems.

In general, the indistinguishability of identical particles requires
interparticle interactions to be permutation invariant. This
permutation invariance places a stringent constraint on possible
many-body interactions. For example, for two-body contact interactions
in one spatial dimension, identical bosons and fermions (with no
internal degrees of freedom) have no other choice but to interact via
delta- and epsilon-function potentials, respectively
\cite{Cheon:1998iy}. In contrast, there is no such constraint for
distinguishable particles, meaning that there are much more variety of
two-body contact interactions for nonidentical particles than those
for identical particles. In fact, in one spatial dimension, the most
general two-body contact interaction consistent with the probability
conservation (unitarity) is known to be described by a $2\times 2$
unitary matrix $U\in U(2)$ \cite{Caudrelier:2005eb,Yonezawa:2005ne},
which contains $\dim U(2)=4$ independent real parameters. However,
scale-invariance breaking under this $U(2)$ family of two-body
interactions has never been studied in the literature.

This paper is aimed at studying breakdown of continuous scale
invariance to discrete scale invariance under the $U(2)$ family of
two-body contact interactions. To simplify the analysis, we shall
focus on two-body problems on the half line, which is exactly solvable
and known to exhibit discrete scale invariance for the case of
identical particles \cite{Ohya:2022yfm}. We shall show that, in a
certain subspace of the parameter space of $U(2)$, there arise two
distinct channels that admit geometric sequences of two-body bound
states.

The rest of the paper is organized as follows. We first formulate the
problem in section \ref{section:2}. We consider two nonidentical
particles on the half line with two-body contact interaction. Under
the assumption that the interaction potential satisfies a scaling law,
we see that the two-body problem is reduced to a set of two
differential equations---the radial and angular Schr\"{o}dinger
equations---where the former determines the two-body energy spectrum
while the latter determines the scale-invariance breaking. Section
\ref{section:3} studies the whole parameter space of scale-invariant
two-body contact interactions in one dimension. We first show that,
for nonidentical particles, there exists a $U(2)$ family of
scale-invariant two-body contact interactions. We then discuss the
quantization condition for eigenvalues of the angular Schr\"{o}dinger
equation. This quantization condition becomes particularly simple if
the masses of the particles are the same. Section \ref{section:4}
studies such a mass-degenerate case and shows that, in a certain
parameter region, there arise two distinct geometric sequences of
two-body bound states. We conclude in section
\ref{section:5}. Appendix \ref{appendix:A} presents a derivation of
the $U(2)$ family of two-body contact interactions.

\section{Scale-invariant two-body problem on the half line}
\label{section:2}
To begin with, let us fix some notation. Let $x_{1}$ and $x_{2}$ be
the coordinates of two nonidentical particles of mass $m_{1}$ and
$m_{2}$ on the half line $\mathbb{R}_{+}=\{x\in\mathbb{R}:x\geq0\}$,
respectively. Let $V(x_{1},x_{2})$ be a potential for interparticle
interactions between these particles. The Hamiltonian of such a
two-body system is then given by
\begin{align}
  H=-\sum_{j=1}^{2}\frac{\hbar^{2}}{2m_{j}}\frac{\partial^{2}}{\partial x_{j}^{2}}+V(x_{1},x_{2}).\label{eq:1}
\end{align}

We wish to understand when and how continuous scale invariance is
broken down to discrete scale invariance under a general two-body
contact interaction. To study this, we focus on the potential $V$ that
fulfills the following properties:
\begin{itemize}
\item\textbf{Property 1. (Scaling law)}
  \begin{align}
    V(\e^{t}x_{1},\e^{t}x_{2})=\e^{-2t}V(x_{1},x_{2}),\quad\forall t\in\mathbb{R}.\label{eq:2}
  \end{align}
\item\textbf{Property 2. (Contactness)}
  \begin{align}
    \supp(V)=\{(x_{1},x_{2})\in\mathbb{R}_{+}^{2}:x_{1}=x_{2}\}.\label{eq:3}
  \end{align}
\end{itemize}
Here $\supp(V)$ stands for the support of $V$. The first property
\eqref{eq:2} is necessary to be scale invariant. The second property
\eqref{eq:3}, on the other hand, describes the situation where the
particles interact only at two-body coincidence points. As we will see
in the next section, such contact interactions are best described by
connection conditions at the two-particle coincidence
points. Throughout this section, however, we shall use the potential
$V$ as a convenient placeholder. Note that we do not impose the
translation invariance $V(x_{1}+a,x_{2}+a)=V(x_{1},x_{2})$
($a\in\mathbb{R}$) on the potential because the translation invariance
is already broken by the boundary.

Now, let us first discuss the impact of the scaling law \eqref{eq:2}
in two-body problems on the half line. To this end, let us introduce
the polar coordinate system $(r,\theta)$ in the two-body configuration
space $\mathbb{R}_{+}^{2}$. We write
\begin{subequations}
  \begin{align}
    x_{1}&=\sqrt{\frac{m_{0}}{m_{1}}}r\cos\theta,\label{eq:4a}\\
    x_{2}&=\sqrt{\frac{m_{0}}{m_{2}}}r\sin\theta,\label{eq:4b}
  \end{align}
\end{subequations}
or, equivalently,
\begin{subequations}
  \begin{align}
    r&=\sqrt{\frac{m_{1}x_{1}^{2}+m_{2}x_{2}^{2}}{m_{0}}},\label{eq:5a}\\
    \theta&=\arctan\left(\sqrt{\frac{m_{2}}{m_{1}}}\frac{x_{2}}{x_{1}}\right),\label{eq:5b}
  \end{align}
\end{subequations}
where $0\leq r<\infty$ and $0\leq\theta\leq\pi/2$. Here $m_{0}(>0)$ is
an arbitrary mass scale introduced to assign the length dimension to
the radius $r$. In this polar coordinate system, the general solution
to the scaling law \eqref{eq:3} can be written as
\begin{align}
  V(x_{1},x_{2})=\frac{\hbar^{2}}{2m_{0}}\frac{v(\theta)}{r^{2}},\label{eq:6}
\end{align}
where $v(\theta)$ is some function of $\theta$. Note that the solution
\eqref{eq:6} is in general not invariant under the translation
$(x_{1},x_{2})\mapsto(x_{1}+a,x_{2}+a)$ and is different from the
translation-invariant two-body non-contact interaction
$V(x_{1},x_{2})\propto1/(x_{1}-x_{2})^{2}$ discussed in the literature
\cite{Moroz:2015}. By using \eqref{eq:6} and the identity
$(1/m_{1})\partial^{2}/\partial
x_{1}^{2}+(1/m_{2})\partial^{2}/\partial
x_{2}^{2}=(1/m_{0})r^{-1/2}(\partial^{2}/\partial
r^{2}+(1/r^{2})\partial^{2}/\partial\theta^{2})r^{1/2}$, we find
\begin{align}
  H=\frac{\hbar^{2}}{2m_{0}}r^{-1/2}\left(-\frac{\partial^{2}}{\partial r^{2}}+\frac{-\partial_{\theta}^{2}+v(\theta)-1/4}{r^{2}}\right)r^{1/2}.\label{eq:7}
\end{align}
The eigenvalue equation of this operator is easily solved by the
method of separation of variables. In fact, by assuming the energy
eigenfunction $\psi$ is of the form,
\begin{align}
  \psi(x_{1},x_{2})=r^{-1/2}R(r)\Theta(\theta),\label{eq:8}
\end{align}
we immediately see that the eigenvalue equation $H\psi=E\psi$ is
reduced to the following set of differential equations:
\begin{subequations}
  \begin{align}
    \left(-\frac{d^{2}}{d\theta^{2}}+v(\theta)\right)\Theta(\theta)&=\lambda\Theta(\theta),\label{eq:9a}\\
    \left(-\frac{d^{2}}{dr^{2}}+\frac{\lambda-1/4}{r^{2}}\right)R(r)&=\frac{2m_{0}E}{\hbar^{2}}R(r).\label{eq:9b}
  \end{align}
\end{subequations}
Notice that the energy eigenvalue $E$ is solely determined by the
one-body problem on the half line with the inverse-square potential
\eqref{eq:9b}, which is known to break continuous scale invariance to
discrete scale invariance when $\lambda<0$ \cite{Case:1950an}. In this
case, there arises a geometric sequence of negative energy eigenvalues
$\{E_{n}\}$ that satisfy the scaling law
$E_{n}/E_{n-1}=\exp(-2\pi/\sqrt{-\lambda})$. Hence, what we have to do
is to solve the angular equation \eqref{eq:9a} and to find out when
the eigenvalue $\lambda$ becomes negative. Let us next turn to this
problem by using the most general scale-invariant two-body contact
interactions consistent with the unitarity.

\section{\texorpdfstring{$U(2)$}{U(2)} family of scale-invariant
  two-body contact interactions}
\label{section:3}
In the previous section, we have imposed the scaling law \eqref{eq:2}
and studied its consequence in the two-body problem on the half
line. In this section, we further impose the contactness \eqref{eq:3},
which, in terms of the function $v(\theta)$, is described by
\begin{align}
  \supp(v)=\{\theta\in[0,\pi/2]:\theta=\theta_{0}\}.\label{eq:10}
\end{align}
Here $\theta_{0}$ is the two-body coincidence point in the polar
coordinate system $(r,\theta)$ defined through \eqref{eq:5b} with
$x_{1}=x_{2}$:
\begin{align}
  \theta_{0}=\arctan\left(\sqrt{\frac{m_{2}}{m_{1}}}\right).\label{eq:11}
\end{align}

Now, we wish to consider the two-body contact interaction as general
as possible. One obvious way to do this is to use the most general
function $v(\theta)$ with one-point support at the two-body
coincidence point $\theta=\theta_{0}$. Another way to do this is to
use the most general connection condition at
$\theta=\theta_{0}$. Since the latter approach is much more
well-established than the former, we shall use the
connection-condition approach to the two-body contact
interaction. Following \cite{Fulop:1999pf}, we shall focus on the most
general connection condition consistent with the probability
conservation, which, in our two-body problem, can be put into the
following continuity condition for the probability current normal to
the two-body coincidence line $x_{1}=x_{2}$ in the two-body
configuration space $\mathbb{R}_{+}^{2}$:
\begin{align}
  \bm{n}\cdot\bm{j}\bigr|_{x_{1}-x_{2}=0_{+}}=\bm{n}\cdot\bm{j}\bigr|_{x_{1}-x_{2}=0_{-}}.\label{eq:12}
\end{align}
Here $\bm{j}=(j_{1},j_{2})$ is the two-body probability current and
$\bm{n}=(n_{1},n_{2})$ is a (unit) normal vector to the line
$x_{1}=x_{2}$. In the original Cartesian coordinate system, they are
given by
$j_{a}=(\hbar/(2im_{a}))(\overline{\psi}\partial\psi/\partial
x_{a}-\overline{\partial\psi/\partial x_{a}}\psi)$ and
$n_{a}=1/\sqrt{2}$, where $a=1,2$ and overline
($\overline{\phantom{m}}$) stands for the complex conjugate. In the
following, we would like to obtain the most general connection
condition for $\Theta(\theta)$ by solving \eqref{eq:12}.

First, in the polar coordinate system, the probability conservation
\eqref{eq:12} at the two-body coincidence point is equivalent to to
the following condition:
\begin{align}
  \overline{\Theta(\theta_{0}+)}\Theta^{\prime}(\theta_{0}+)-\overline{\Theta^{\prime}(\theta_{0}+)}\Theta(\theta_{0}+)=\overline{\Theta(\theta_{0}-)}\Theta^{\prime}(\theta_{0}-)-\overline{\Theta^{\prime}(\theta_{0}-)}\Theta(\theta_{0}-),\label{eq:13}
\end{align}
where prime (${}^{\prime}$) indicates the derivative with respect to
$\theta$ and $f(\theta_{0}\pm)$ stands for one-sided limits given by
$f(\theta_{0}\pm)=\lim_{\epsilon\to0_{+}}f(\theta_{0}\pm\epsilon)$. Notice
that \eqref{eq:13} is quadratic equation with respect to
$\Theta$. However, it can be linearized and there exists a $U(2)$
family of linearized solutions. An important observation to see this
is that \eqref{eq:13} can be put into the form of inner product in
two-dimensional complex vector space $\mathbb{C}^{2}$:
\begin{align}
  \begin{pmatrix}\Theta(\theta_{0}+)\\\Theta(\theta_{0}-)\end{pmatrix}^{\dagger}\begin{pmatrix}\Theta^{\prime}(\theta_{0}+)\\-\Theta^{\prime}(\theta_{0}-)\end{pmatrix}
  =\begin{pmatrix}\Theta^{\prime}(\theta_{0}+)\\-\Theta^{\prime}(\theta_{0}-)\end{pmatrix}^{\dagger}\begin{pmatrix}\Theta(\theta_{0}+)\\\Theta(\theta_{0}-)\end{pmatrix},\label{eq:14}
\end{align}
where ${}^{\dagger}$ stands for the hermitian conjugate. As shown in
appendix \ref{appendix:A}, the solution to this type of equation is
parameterized by a $2\times2$ unitary matrix $U$ and given by
\begin{align}
  (1-U)\begin{pmatrix}\Theta(\theta_{0}+)\\\Theta(\theta_{0}-)\end{pmatrix}+i(1+U)\begin{pmatrix}\Theta^{\prime}(\theta_{0}+)\\-\Theta^{\prime}(\theta_{0}-)\end{pmatrix}=0,\quad U\in U(2).\label{eq:15}
\end{align}
This is the $U(2)$ family of connection conditions that describe all
possible scale-invariant two-body contact interactions in the two-body
problem on the half line.

Now, in order to solve the problem, we also have to specify the
boundary conditions at $\theta=0,\pi/2$. For simplicity, we will
impose the Dirichlet boundary conditions:
\begin{align}
  \Theta(0)=\Theta(\pi/2)=0.\label{eq:16}
\end{align}
What we have to do is then to solve the angular equation
$-\Theta^{\prime\prime}(\theta)=\lambda\Theta(\theta)$ for
$\theta\neq\theta_{0}$ under the conditions \eqref{eq:15} and
\eqref{eq:16} and to determine the unitary matrix $U$ that admits
negative eigenvalue $\lambda$.

To this end, let us next parameterize the unitary matrix $U$. Let
$\e^{i\alpha_{+}}$ and $\e^{i\alpha_{-}}$ be two eigenvalues of $U$,
where $\alpha_{\pm}\in[-\pi,\pi)$ are real parameters. Then, the
$2\times2$ unitary matrix $U$ can be written as the following spectral
decomposition:
\begin{align}
  U=\e^{i\alpha_{+}}P_{+}+\e^{i\alpha_{-}}P_{-},\label{eq:17}
\end{align}
where $P_{\pm}$ are projection operators that satisfy the completeness
$P_{+}+P_{-}=1$, hermiticity $P_{\pm}^{\dagger}=P_{\pm}$, and
orthonormality $P_{a}P_{b}=\delta_{ab}P_{b}$ ($a,b\in\{+,-\}$). Such
$2\times2$ hermitian matrices are parameterized as follows:
\begin{align}
  P_{\pm}=\frac{1\pm\bm{e}\cdot\bm{\sigma}}{2},\label{eq:18}
\end{align}
where $\bm{e}=(e_{1},e_{2},e_{3})$ is a real unit 3-vector satisfying
$e_{1}^{2}+e_{2}^{2}+e_{3}^{2}=1$ and
$\bm{\sigma}=(\sigma_{1},\sigma_{2},\sigma_{3})$ is the 3-tuple of
Pauli matrices. Eq.~\eqref{eq:17} is the most general parameterization
of $U\in U(2)$, which contains four independent parameters
$(\alpha_{+},\alpha_{-},\bm{e})$. What we want to do is to determine
the parameter space of $U$ that admits negative eigenvalue
$\lambda$. Let us next do this by solving the angular equation
$-\Theta^{\prime\prime}(\theta)=\lambda\Theta(\theta)$.

First, the general solution to the angular equation with the Dirichlet
boundary conditions \eqref{eq:16} can be written as the following
linear combination of plane waves:
\begin{align}
  \Theta(\theta)=
  \begin{cases}
    A(\e^{i\sqrt{\lambda}(\theta_{0}-\theta)}-\e^{i\sqrt{\lambda}(\theta_{0}+\theta-\pi)})&\text{for $\theta\in(\theta_{0},\pi/2]$},\\
    B(\e^{i\sqrt{\lambda}(\theta-\theta_{0})}-\e^{-i\sqrt{\lambda}(\theta+\theta_{0})})&\text{for $\theta\in[0,\theta_{0})$}.\\
  \end{cases}\label{eq:19}
\end{align}
Here $A$ and $B$ are integration constants. Next, we impose the
connection condition \eqref{eq:15}. Substituting \eqref{eq:19} into
\eqref{eq:15} we get the following condition:
\begin{align}
  (1-\mathscr{U}(\lambda))\begin{pmatrix}A\\B\\\end{pmatrix}=0,\label{eq:20}
\end{align}
where
\begin{align}
  \mathscr{U}(\lambda)=\sum_{j=\pm}\e^{2i[\frac{\pi}{4}\sqrt{\lambda}+\arctan(\sqrt{\lambda}\cot(\alpha_{j}/2))]}P_{j}\e^{2i\sqrt{\lambda}(\frac{\pi}{4}-\theta_{0})\sigma_{3}}.\label{eq:21}
\end{align}
Hence, in order to have nontrivial solution $(A,B)\neq(0,0)$, the
matrix $1-\mathscr{U}(\lambda)$ must not have the inverse for any
$\lambda$. Thus we get
\begin{align}
  \det(1-\mathscr{U}(\lambda))=0.\label{eq:22}
\end{align}
This gives the quantization condition for the eigenvalue
$\lambda$. This equation possesses negative solutions in a certain
parameter region of $U$. Let us next see this by considering the
simple case $m_{1}=m_{2}$.

\section{A simple example: mass-degenerate case}
\label{section:4}
\begin{figure}[t]
  \centering \input{figure1.eepic}
  \caption{The $\alpha_{\pm}$-dependence of the eigenvalues
    $\lambda_{\pm,0}<\lambda_{\pm,1}<\lambda_{\pm,2}<\cdots$. The red
    dot represents the critical value $\alpha=\alpha_{\ast}$ below
    which the lowest eigenvalue $\lambda_{\pm,0}$ becomes negative.}
  \label{figure:1}
\end{figure}
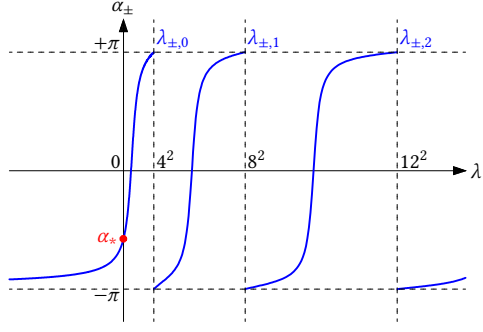

The quantization condition \eqref{eq:22} becomes particularly simple
in the mass-degenerate case $m_{1}=m_{2}$, where the angle
$\theta_{0}$ becomes
\begin{align}
  \theta_{0}=\arctan(1)=\frac{\pi}{4}\quad\text{for $m_{1}=m_{2}$}.\label{eq:23}
\end{align}
In this case, the matrix \eqref{eq:21} takes the following form:
\begin{align}
  \mathscr{U}(\lambda)=\sum_{j=\pm}\e^{2i[\frac{\pi}{4}\sqrt{\lambda}+\arctan(\sqrt{\lambda}\cot(\alpha_{j}/2))]}P_{j}.\label{eq:24}
\end{align}
It is clear from this expression that the unitary matrix
$\mathscr{U}(\lambda)$ has the eigenvalues
$\e^{2i[\frac{\pi}{4}\sqrt{\lambda}+\arctan(\sqrt{\lambda}\cot(\alpha_{\pm}/2))]}$,
which must be $1$ in order to fulfill the condition
\eqref{eq:22}. Thus we find the following quantization condition:
\begin{align}
  \frac{\pi}{4}\sqrt{\lambda}+\arctan\left(\sqrt{\lambda}\cot\left(\frac{\alpha_{\pm}}{2}\right)\right)=n\pi,\quad(n:\text{integer}),\label{eq:25}
\end{align}
or, equivalently,
\begin{align}
  \alpha_{\pm}=-2\arctan\left(\sqrt{\lambda}\cot\left(\frac{\pi}{4}\sqrt{\lambda}\right)\right).\label{eq:26}
\end{align}
This equation is easily solved graphically and possesses infinitely
many real roots
$\lambda_{\pm,0}<\lambda_{\pm,1}<\lambda_{\pm,2}<\cdots$. As depicted
in Fig.~\ref{figure:1}, the lowest root $\lambda_{\pm,0}$ becomes
negative for $\alpha_{\pm}<\alpha_{\ast}$, where $\alpha_{\ast}$ is
the critical value given by
\begin{align}
  \alpha_{\ast}=\lim_{\lambda\to0}\left[-2\arctan\left(\sqrt{\lambda}\cot\left(\frac{\pi}{4}\sqrt{\lambda}\right)\right)\right]=-2\arctan\left(\frac{4}{\pi}\right).\label{eq:27}
\end{align}
Hence, continuous scale invariance is broken to discrete scale
invariance for $\alpha_{\pm}<\alpha_{\ast}$. In consequence, there
exist three distinct ``phases'' in this model. The first is the region
$\{(\alpha_{+},\alpha_{-}):\alpha_{+}<\alpha_{\ast}~~\&~~\alpha_{-}<\alpha_{\ast}\}$
in which there arise two distinct geometric series of two-body bound
states in the $\lambda_{+,0}$- and $\lambda_{-,0}$-channels; the
second is the region
$\{(\alpha_{+},\alpha_{-}):\alpha_{\pm}<\alpha_{\ast}~~\&~~\alpha_{\mp}\geq\alpha_{\ast}\}$
in which there arises a single geometric series of two-body bound
states in the $\lambda_{\pm,0}$-channel; and the third is the region
$\{(\alpha_{+},\alpha_{-}):\alpha_{+}\geq\alpha_{\ast}~~\&~~\alpha_{+}\geq\alpha_{\ast}\}$
in which continuous scale invariance is unbroken and there is no
two-body bound states. Notice that the parameter $\bm{e}$ has no
effect in this phase structure.

\section{Conclusion}
\label{section:5}
In this paper, we have studied breakdown of continuous scale
invariance to discrete invariance in two-body problems of nonidentical
particles on the half line. We first discussed that, for nonidentical
particles in one dimension, there exists the $U(2)$ family of
scale-invariant two-body contact interactions consistent with the
probability conservation. We then discussed the breakdown of
continuous scale invariance under this $U(2)$ family of
interactions. We showed that, in addition to the unbroken phase, there
exist two additional phases in the scale-invariant two-body problem:
one is the discrete scale invariant phase in which two distinct
geometric series of two-body bound states appear, and the other is the
discrete scale invariant phase in which a single geometric series of
two-body bound states appear. This is in sharp contrast to the
two-body problem of identical particles on the half line
\cite{Ohya:2022yfm}, where there arises at most a single geometric
series of two-body bound states.

Throughout this paper we focused on two-body problems with boundary
just for simplicity. Future studies should investigate a
generalization to $n(\geq3)$-body problems of nonidentical particles
with or without boundary and reveal its phase structure. It is also
interesting to study experimental realizations of the scenario
discussed in this paper. Though currently unknown, the realization of
the $U(2)$ family of two-body contact interactions in experiments will
open up a new phase of many-body problems in one dimension.

\subsection*{Acknowledgment}
This work was supported by JSPS KAKENHI Grant Number JP23K03267.

\begin{appendices}
  \section{Proof of \texorpdfstring{\eqref{eq:15}}{(15)}}
  \label{appendix:A}
  Following the original work by F\"{u}l\"{o}p and Tsutsui
  \cite{Fulop:1999pf}, in this section we show that there exists a
  $U(2)$ family of solutions to the continuity condition
  \eqref{eq:14}. The essentially same derivation can also be found in
  \cite{Bonneau:1999zq}.

  To begin with, let us consider the following abstract problem. Let
  $\Psi$ and $\Phi$ be some elements of an arbitrary Hilbert space
  $\mathscr{H}$ and subject to the following condition:
  \begin{align}
    (\Psi,\Phi)=(\Phi,\Psi),\label{eq:A1}
  \end{align}
  where $(\ast,\ast)$ stands for the inner product. We wish to find
  out all possible solutions to this equation. To solve this, we first
  note that any inner products can be written in terms of norms. In
  fact, the polarization identity gives (see, e.g., Eq.~(1.11) of
  \cite{Amrein:2009})
  \begin{align}
    (\Psi,\Phi)=\frac{1}{4}\left(\|\Psi+\Phi\|^{2}-\|\Psi-\Phi\|^{2}+i\|\Psi+i\Phi\|^{2}-i\|\Psi-i\Phi\|^{2}\right),\label{eq:A2}
  \end{align}
  where $\|\ast\|$ stands for the norm. Note that
  $(\Phi,\Psi)=\overline{(\Psi,\Phi)}=(1/4)(\|\Psi+\Phi\|^{2}-\|\Psi-\Phi\|^{2}-i\|\Psi+i\Phi\|^{2}+i\|\Psi-i\Phi\|^{2})$. Equating
  these expressions, one immediately sees that \eqref{eq:A1} is
  equivalent to the following condition:
  \begin{align}
    \|\Psi+i\Phi\|^{2}=\|\Psi-i\Phi\|^{2}.\label{eq:A3}
  \end{align}
  This equation says that two distinct vectors $\Psi+i\Phi$ and
  $\Psi-i\Phi$ have the same norm. Hence there should exist a unitary
  $U$ that satisfies the following relation:
  \begin{align}
    \Psi+i\Phi=U(\Psi-i\Phi),\label{eq:A4}
  \end{align}
  or, equivalently,
  \begin{align}
    (1-U)\Psi+i(1+U)\Phi=0.\label{eq:A5}
  \end{align}
  This is the linearized relation that solves \eqref{eq:A1}.

  Now let us turn to the problem of solving the continuity condition
  \eqref{eq:14}. We first note that \eqref{eq:14} is equivalent to
  \eqref{eq:A1} with the special choice $\mathscr{H}=\mathbb{C}^{2}$,
  $\Psi={}^{t}(\Theta(\theta_{0}+),\Theta(\theta_{0}-))$,
  $\Phi={}^{t}(\Theta^{\prime}(\theta_{0}+),-\Theta^{\prime}(\theta_{0}-))$,
  and $(\Psi,\Phi)=\Psi^{\dagger}\Phi$, where ${}^{t}$ stands for the
  transposition. Note also that $U$ in this case is just a $2\times2$
  unitary matrix. Hence there exists a $U(2)$ family of connection
  conditions given by \eqref{eq:15}. This completes the proof.
\end{appendices}

\printbibliography[heading=bibintoc]
\end{document}

%% file: figure1.eepic
\xdefinecolor{rgb_000000}{rgb}{0,0,0}%
\xdefinecolor{rgb_0000ff}{rgb}{0,0,1}%
\xdefinecolor{rgb_ff0000}{rgb}{1,0,0}%
\setlength{\unitlength}{1cm}%
\begin{picture}(6,4)(0,0)%
\path(0,2)(6,2)
\path(5.84184,1.96046)(5.88138,1.96046)
\path(5.84184,1.97364)(5.92092,1.97364)
\path(5.84184,1.98682)(5.96046,1.98682)
\path(5.84184,2)(6,2)
\path(5.84184,2.01318)(5.96046,2.01318)
\path(5.84184,2.02636)(5.92092,2.02636)
\path(5.84184,2.03954)(5.88138,2.03954)
\path(5.98682,1.99561)(5.98682,2.00439)
\path(5.97364,1.99121)(5.97364,2.00879)
\path(5.96046,1.98682)(5.96046,2.01318)
\path(5.94728,1.98243)(5.94728,2.01757)
\path(5.9341,1.97803)(5.9341,2.02197)
\path(5.92092,1.97364)(5.92092,2.02636)
\path(5.90774,1.96925)(5.90774,2.03075)
\path(5.89456,1.96485)(5.89456,2.03515)
\path(5.88138,1.96046)(5.88138,2.03954)
\path(5.8682,1.95607)(5.8682,2.04393)
\path(5.85502,1.95167)(5.85502,2.04833)
\path(5.84184,1.94728)(5.84184,2.05272)
\path(5.84184,2)(5.84184,1.94728)(6,2)(5.84184,2.05272)(5.84184,2)
\path(1.5,0)(1.5,4)
\path(1.54722,3.84184)(1.49341,3.98023)
\path(1.53294,3.84184)(1.48682,3.96046)
\path(1.51866,3.84184)(1.48023,3.94069)
\path(1.50439,3.84184)(1.47364,3.92092)
\path(1.49011,3.84184)(1.46705,3.90115)
\path(1.47583,3.84184)(1.46046,3.88138)
\path(1.46156,3.84184)(1.45387,3.86161)
\path(1.50439,3.98682)(1.4943,3.9829)
\path(1.50879,3.97364)(1.4886,3.96579)
\path(1.51318,3.96046)(1.4829,3.94869)
\path(1.51757,3.94728)(1.47719,3.93158)
\path(1.52197,3.9341)(1.47149,3.91448)
\path(1.52636,3.92092)(1.46579,3.89737)
\path(1.53075,3.90774)(1.46009,3.88027)
\path(1.53515,3.89456)(1.45439,3.86316)
\path(1.53954,3.88138)(1.44869,3.84606)
\path(1.54393,3.8682)(1.47614,3.84184)
\path(1.54833,3.85502)(1.51443,3.84184)
\path(1.5,3.84184)(1.55272,3.84184)(1.5,4)(1.44728,3.84184)(1.5,3.84184)
\allinethickness{0.8pt}%
\color{rgb_0000ff}%
\path(0,0.557592)(0.005,0.557804)(0.01,0.558017)(0.015,0.558231)
  (0.02,0.558447)(0.025,0.558663)(0.03,0.55888)(0.035,0.559099)
  (0.04,0.559319)(0.045,0.55954)(0.05,0.559762)(0.055,0.559985)
  (0.06,0.560209)(0.065,0.560434)(0.07,0.560661)(0.075,0.560889)
  (0.08,0.561117)(0.085,0.561348)(0.09,0.561579)(0.095,0.561811)
  (0.1,0.562045)(0.105,0.56228)(0.11,0.562517)(0.115,0.562754)
  (0.12,0.562993)(0.125,0.563233)(0.13,0.563474)(0.135,0.563717)
  (0.14,0.563961)(0.145,0.564206)(0.15,0.564453)(0.155,0.564701)
  (0.16,0.56495)(0.165,0.565201)(0.17,0.565453)(0.175,0.565707)
  (0.18,0.565962)(0.185,0.566219)(0.19,0.566476)(0.195,0.566736)
  (0.2,0.566997)(0.205,0.567259)(0.21,0.567523)(0.215,0.567788)
  (0.22,0.568055)(0.225,0.568323)(0.23,0.568593)(0.235,0.568865)
  (0.24,0.569138)(0.245,0.569413)(0.25,0.569689)(0.255,0.569967)
  (0.26,0.570247)(0.265,0.570528)(0.27,0.570811)(0.275,0.571096)
  (0.28,0.571382)(0.285,0.571671)(0.29,0.571961)(0.295,0.572252)
  (0.3,0.572546)(0.305,0.572841)(0.31,0.573138)(0.315,0.573437)
  (0.32,0.573738)(0.325,0.574041)(0.33,0.574346)(0.335,0.574652)
  (0.34,0.574961)(0.345,0.575272)(0.35,0.575584)(0.355,0.575899)
  (0.36,0.576215)(0.365,0.576534)(0.37,0.576855)(0.375,0.577177)
  (0.38,0.577502)(0.385,0.577829)(0.39,0.578159)(0.395,0.57849)
  (0.4,0.578824)(0.405,0.57916)(0.41,0.579498)(0.415,0.579838)
  (0.42,0.580181)(0.425,0.580526)(0.43,0.580874)(0.435,0.581224)
  (0.44,0.581576)(0.445,0.581931)(0.45,0.582288)(0.455,0.582648)
  (0.46,0.58301)(0.465,0.583375)(0.47,0.583743)(0.475,0.584113)
  (0.48,0.584486)(0.485,0.584861)(0.49,0.58524)(0.495,0.585621)
  (0.5,0.586004)(0.505,0.586391)(0.51,0.586781)(0.515,0.587173)
  (0.52,0.587568)(0.525,0.587967)(0.53,0.588368)(0.535,0.588773)
  (0.54,0.58918)(0.545,0.589591)(0.55,0.590004)(0.555,0.590421)
  (0.56,0.590842)(0.565,0.591265)(0.57,0.591692)(0.575,0.592122)
  (0.58,0.592556)(0.585,0.592993)(0.59,0.593434)(0.595,0.593878)
  (0.6,0.594326)(0.605,0.594778)(0.61,0.595233)(0.615,0.595692)
  (0.62,0.596155)(0.625,0.596621)(0.63,0.597092)(0.635,0.597567)
  (0.64,0.598045)(0.645,0.598528)(0.65,0.599015)(0.655,0.599506)
  (0.66,0.600001)(0.665,0.600501)(0.67,0.601005)(0.675,0.601514)
  (0.68,0.602027)(0.685,0.602544)(0.69,0.603066)(0.695,0.603593)
  (0.7,0.604125)(0.705,0.604662)(0.71,0.605204)(0.715,0.60575)
  (0.72,0.606302)(0.725,0.606859)(0.73,0.607421)(0.735,0.607989)
  (0.74,0.608562)(0.745,0.60914)(0.75,0.609724)(0.755,0.610314)
  (0.76,0.610909)(0.765,0.611511)(0.77,0.612118)(0.775,0.612732)
  (0.78,0.613351)(0.785,0.613977)(0.79,0.614609)(0.795,0.615248)
  (0.8,0.615894)(0.805,0.616546)(0.81,0.617205)(0.815,0.61787)
  (0.82,0.618543)(0.825,0.619223)(0.83,0.619911)(0.835,0.620606)
  (0.84,0.621308)(0.845,0.622018)(0.85,0.622737)(0.855,0.623463)
  (0.86,0.624197)(0.865,0.624939)(0.87,0.62569)(0.875,0.62645)
  (0.88,0.627218)(0.885,0.627996)(0.89,0.628782)(0.895,0.629578)
  (0.9,0.630383)(0.905,0.631198)(0.91,0.632023)(0.915,0.632858)
  (0.92,0.633703)(0.925,0.634559)(0.93,0.635425)(0.935,0.636303)
  (0.94,0.637191)(0.945,0.638091)(0.95,0.639002)(0.955,0.639925)
  (0.96,0.640861)(0.965,0.641808)(0.97,0.642769)(0.975,0.643742)
  (0.98,0.644729)(0.985,0.645729)(0.99,0.646743)(0.995,0.647771)
  (1,0.648813)(1.005,0.64987)(1.01,0.650943)(1.015,0.652031)
  (1.02,0.653134)(1.025,0.654254)(1.03,0.655391)(1.035,0.656545)
  (1.04,0.657716)(1.045,0.658905)(1.05,0.660112)(1.055,0.661338)
  (1.06,0.662584)(1.065,0.663849)(1.07,0.665134)(1.075,0.666441)
  (1.08,0.667768)(1.085,0.669118)(1.09,0.67049)(1.095,0.671886)
  (1.1,0.673305)(1.105,0.674749)(1.11,0.676218)(1.115,0.677712)
  (1.12,0.679234)(1.125,0.680782)(1.13,0.682359)(1.135,0.683965)
  (1.14,0.685601)(1.145,0.687268)(1.15,0.688967)(1.155,0.690698)
  (1.16,0.692463)(1.165,0.694263)(1.17,0.696098)(1.175,0.697971)
  (1.18,0.699883)(1.185,0.701834)(1.19,0.703825)(1.195,0.70586)
  (1.2,0.707938)(1.205,0.710061)(1.21,0.712231)(1.215,0.71445)
  (1.22,0.716719)(1.225,0.71904)(1.23,0.721416)(1.235,0.723847)
  (1.24,0.726337)(1.245,0.728888)(1.25,0.731501)
\path(1.25,0.731501)(1.255,0.73418)(1.26,0.736926)(1.265,0.739744)
  (1.27,0.742635)(1.275,0.745603)(1.28,0.748651)(1.285,0.751782)
  (1.29,0.755001)(1.295,0.75831)(1.3,0.761715)(1.305,0.765219)
  (1.31,0.768827)(1.315,0.772544)(1.32,0.776375)(1.325,0.780326)
  (1.33,0.784403)(1.335,0.788611)(1.34,0.792958)(1.345,0.797451)
  (1.35,0.802098)(1.355,0.806906)(1.36,0.811885)(1.365,0.817044)
  (1.37,0.822394)(1.375,0.827945)(1.38,0.833709)(1.385,0.839698)
  (1.39,0.845928)(1.395,0.852412)(1.4,0.859167)(1.405,0.866209)
  (1.41,0.873559)(1.415,0.881236)(1.42,0.889264)(1.425,0.897666)
  (1.43,0.906469)(1.435,0.915702)(1.44,0.925398)(1.445,0.935591)
  (1.45,0.946319)(1.455,0.957626)(1.46,0.969557)(1.465,0.982165)
  (1.47,0.995505)(1.475,1.00964)(1.48,1.02464)(1.485,1.04059)
  (1.49,1.05756)(1.495,1.07565)
\path(1.50133,1.10035)(1.50267,1.10583)(1.504,1.1114)
  (1.50533,1.11707)(1.50667,1.12286)(1.508,1.12875)(1.50933,1.13475)
  (1.51067,1.14086)(1.512,1.14709)(1.51333,1.15344)(1.51467,1.15992)
  (1.516,1.16651)(1.51733,1.17324)(1.51867,1.1801)(1.52,1.1871)
  (1.52133,1.19423)(1.52267,1.20151)(1.524,1.20893)(1.52533,1.2165)
  (1.52666,1.22423)(1.528,1.23211)(1.52933,1.24015)(1.53066,1.24836)
  (1.532,1.25674)(1.53333,1.2653)(1.53466,1.27403)(1.536,1.28294)
  (1.53733,1.29204)(1.53866,1.30133)(1.54,1.31082)(1.54133,1.32051)
  (1.54266,1.33041)(1.544,1.34051)(1.54533,1.35083)(1.54666,1.36137)
  (1.548,1.37213)(1.54933,1.38312)(1.55066,1.39435)(1.552,1.40582)
  (1.55333,1.41753)(1.55466,1.42949)(1.556,1.4417)(1.55733,1.45417)
  (1.55866,1.4669)(1.56,1.4799)(1.56133,1.49317)(1.56266,1.50672)
  (1.564,1.52054)(1.56533,1.53465)(1.56666,1.54904)(1.568,1.56371)
  (1.56933,1.57868)(1.57066,1.59394)(1.572,1.6095)(1.57333,1.62535)
  (1.57466,1.6415)(1.576,1.65794)(1.57733,1.67467)(1.57866,1.6917)
  (1.57999,1.70903)(1.58133,1.72664)(1.58266,1.74454)
  (1.58399,1.76271)(1.58533,1.78117)(1.58666,1.79989)
  (1.58799,1.81887)(1.58933,1.83811)(1.59066,1.8576)(1.59199,1.87732)
  (1.59333,1.89726)(1.59466,1.91742)(1.59599,1.93778)
  (1.59733,1.95832)(1.59866,1.97903)(1.59999,1.9999)(1.60133,2.02091)
  (1.60266,2.04204)(1.60399,2.06328)(1.60533,2.0846)(1.60666,2.106)
  (1.60799,2.12744)(1.60933,2.14891)(1.61066,2.1704)(1.61199,2.19187)
  (1.61333,2.21332)(1.61466,2.23472)(1.61599,2.25606)
  (1.61733,2.27731)(1.61866,2.29846)(1.61999,2.31949)
  (1.62133,2.34038)(1.62266,2.36112)(1.62399,2.38169)
  (1.62533,2.40208)(1.62666,2.42227)(1.62799,2.44225)
  (1.62933,2.46201)(1.63066,2.48153)(1.63199,2.50081)
  (1.63332,2.51984)(1.63466,2.53861)(1.63599,2.55711)
  (1.63732,2.57533)(1.63866,2.59328)(1.63999,2.61094)
  (1.64132,2.62832)(1.64266,2.6454)(1.64399,2.66219)(1.64532,2.67869)
  (1.64666,2.69489)(1.64799,2.7108)(1.64932,2.72642)(1.65066,2.74174)
  (1.65199,2.75677)(1.65332,2.77151)(1.65466,2.78596)
  (1.65599,2.80013)(1.65732,2.81401)(1.65866,2.82762)
  (1.65999,2.84095)(1.66132,2.85402)(1.66266,2.86682)
  (1.66399,2.87935)(1.66532,2.89163)(1.66666,2.90366)
  (1.66799,2.91543)(1.66932,2.92697)(1.67066,2.93826)
  (1.67199,2.94933)(1.67332,2.96016)(1.67466,2.97077)
  (1.67599,2.98116)(1.67732,2.99133)(1.67866,3.0013)(1.67999,3.01106)
  (1.68132,3.02062)(1.68266,3.02998)(1.68399,3.03916)
  (1.68532,3.04814)(1.68665,3.05695)(1.68799,3.06558)
  (1.68932,3.07403)(1.69065,3.08232)(1.69199,3.09044)(1.69332,3.0984)
  (1.69465,3.1062)(1.69599,3.11385)(1.69732,3.12135)(1.69865,3.12871)
  (1.69999,3.13592)(1.70132,3.143)(1.70265,3.14994)(1.70399,3.15675)
  (1.70532,3.16343)(1.70665,3.16998)(1.70799,3.17642)
  (1.70932,3.18273)(1.71065,3.18893)(1.71199,3.19502)(1.71332,3.201)
  (1.71465,3.20687)(1.71599,3.21263)(1.71732,3.21829)
  (1.71865,3.22386)(1.71999,3.22932)(1.72132,3.23469)
  (1.72265,3.23997)(1.72399,3.24516)(1.72532,3.25026)
  (1.72665,3.25527)(1.72799,3.2602)(1.72932,3.26505)(1.73065,3.26982)
  (1.73199,3.27451)(1.73332,3.27913)(1.73465,3.28367)
  (1.73599,3.28813)(1.73732,3.29253)(1.73865,3.29686)
  (1.73998,3.30112)(1.74132,3.30531)(1.74265,3.30944)
  (1.74398,3.31351)(1.74532,3.31751)(1.74665,3.32146)
  (1.74798,3.32534)(1.74932,3.32917)(1.75065,3.33294)
  (1.75198,3.33666)(1.75332,3.34032)(1.75465,3.34393)
  (1.75598,3.34749)(1.75732,3.351)(1.75865,3.35446)(1.75998,3.35788)
  (1.76132,3.36124)(1.76265,3.36456)(1.76398,3.36784)
  (1.76532,3.37107)(1.76665,3.37425)(1.76798,3.3774)(1.76932,3.3805)
  (1.77065,3.38357)(1.77198,3.38659)(1.77332,3.38957)
  (1.77465,3.39252)(1.77598,3.39543)(1.77732,3.39831)
  (1.77865,3.40114)(1.77998,3.40395)(1.78132,3.40672)
  (1.78265,3.40945)(1.78398,3.41215)(1.78532,3.41482)
  (1.78665,3.41746)(1.78798,3.42007)(1.78932,3.42265)(1.79065,3.4252)
  (1.79198,3.42772)(1.79331,3.43021)(1.79465,3.43267)
  (1.79598,3.43511)(1.79731,3.43751)(1.79865,3.4399)(1.79998,3.44225)
  (1.80131,3.44458)(1.80265,3.44689)(1.80398,3.44917)
  (1.80531,3.45143)(1.80665,3.45366)(1.80798,3.45587)
  (1.80931,3.45806)(1.81065,3.46023)(1.81198,3.46237)
  (1.81331,3.46449)(1.81465,3.4666)(1.81598,3.46868)(1.81731,3.47074)
  (1.81865,3.47278)(1.81998,3.4748)(1.82131,3.47681)(1.82265,3.47879)
  (1.82398,3.48076)(1.82531,3.4827)(1.82665,3.48463)(1.82798,3.48654)
  (1.82931,3.48844)(1.83065,3.49032)(1.83198,3.49218)(1.83331,3.49403)
\path(1.83331,3.49403)(1.83465,3.49586)(1.83598,3.49767)
  (1.83731,3.49947)(1.83865,3.50125)(1.83998,3.50302)
  (1.84131,3.50478)(1.84265,3.50652)(1.84398,3.50824)
  (1.84531,3.50995)(1.84664,3.51165)(1.84798,3.51334)
  (1.84931,3.51501)(1.85064,3.51667)(1.85198,3.51831)
  (1.85331,3.51995)(1.85464,3.52157)(1.85598,3.52318)
  (1.85731,3.52478)(1.85864,3.52636)(1.85998,3.52794)(1.86131,3.5295)
  (1.86264,3.53106)(1.86398,3.5326)(1.86531,3.53413)(1.86664,3.53565)
  (1.86798,3.53716)(1.86931,3.53866)(1.87064,3.54015)
  (1.87198,3.54164)(1.87331,3.54311)(1.87464,3.54457)
  (1.87598,3.54602)(1.87731,3.54747)(1.87864,3.5489)(1.87998,3.55033)
  (1.88131,3.55175)(1.88264,3.55315)(1.88398,3.55456)
  (1.88531,3.55595)(1.88664,3.55733)(1.88798,3.55871)
  (1.88931,3.56008)(1.89064,3.56144)(1.89198,3.5628)(1.89331,3.56414)
  (1.89464,3.56548)(1.89598,3.56681)(1.89731,3.56814)
  (1.89864,3.56946)(1.89997,3.57077)
\path(1.90002,0.429228)(1.91202,0.440757)(1.92402,0.451839)
  (1.93602,0.462548)(1.94802,0.472953)(1.96002,0.483114)
  (1.97202,0.493089)(1.98402,0.502933)(1.99602,0.512696)
  (2.00802,0.522429)(2.02002,0.53218)(2.03202,0.542001)
  (2.04402,0.551941)(2.05602,0.562053)(2.06802,0.572393)
  (2.08002,0.58302)(2.09202,0.593999)(2.10402,0.605401)
  (2.11602,0.617305)(2.12802,0.6298)(2.14001,0.642988)
  (2.15201,0.656986)(2.16401,0.671929)(2.17601,0.687977)
  (2.18801,0.70532)(2.20001,0.724183)(2.21201,0.744838)
  (2.22401,0.767618)(2.23601,0.79293)(2.24801,0.821281)
  (2.26001,0.853302)(2.27201,0.889792)(2.28401,0.931763)
  (2.29601,0.980508)(2.30801,1.03767)(2.32001,1.10534)
  (2.33201,1.18609)(2.34401,1.28294)(2.35601,1.39907)
  (2.36801,1.53698)(2.38,1.69702)(2.392,1.87532)(2.404,2.06288)
  (2.416,2.24746)(2.428,2.41787)(2.44,2.56752)(2.452,2.69473)
  (2.464,2.80112)(2.476,2.88964)(2.488,2.96345)(2.5,3.0254)
  (2.512,3.07781)(2.524,3.12255)(2.536,3.1611)(2.548,3.19459)
  (2.56,3.22394)(2.572,3.24985)(2.584,3.27289)(2.596,3.29351)
  (2.608,3.31208)(2.61999,3.3289)(2.63199,3.34421)(2.64399,3.35821)
  (2.65599,3.37108)(2.66799,3.38295)(2.67999,3.39393)
  (2.69199,3.40415)(2.70399,3.41367)(2.71599,3.42258)
  (2.72799,3.43093)(2.73999,3.4388)(2.75199,3.44621)(2.76399,3.45322)
  (2.77599,3.45987)(2.78799,3.46618)(2.79999,3.47219)
  (2.81199,3.47793)(2.82399,3.48341)(2.83599,3.48866)
  (2.84799,3.49369)(2.85998,3.49853)(2.87198,3.50319)
  (2.88398,3.50769)(2.89598,3.51202)(2.90798,3.51622)
  (2.91998,3.52029)(2.93198,3.52423)(2.94398,3.52806)
  (2.95598,3.53179)(2.96798,3.53542)(2.97998,3.53897)
  (2.99198,3.54243)(3.00398,3.54581)(3.01598,3.54913)
  (3.02798,3.55238)(3.03998,3.55557)(3.05198,3.5587)(3.06398,3.56179)
  (3.07598,3.56483)(3.08798,3.56783)(3.09997,3.57079)
\path(3.10002,0.42921)(3.12002,0.434075)(3.14002,0.438866)
  (3.16002,0.443598)(3.18002,0.448288)(3.20002,0.452951)
  (3.22002,0.457603)(3.24002,0.462258)(3.26002,0.466932)
  (3.28002,0.471641)(3.30002,0.4764)(3.32002,0.481228)
  (3.34002,0.486142)(3.36002,0.491161)(3.38002,0.496306)
  (3.40002,0.501599)(3.42002,0.507066)(3.44002,0.512734)
  (3.46002,0.518634)(3.48002,0.524801)(3.50001,0.531274)
  (3.52001,0.538099)(3.54001,0.545328)(3.56001,0.553024)
  (3.58001,0.561257)(3.60001,0.570113)(3.62001,0.579695)
  (3.64001,0.590125)(3.66001,0.601554)(3.68001,0.614167)
  (3.70001,0.628194)(3.72001,0.643927)(3.74001,0.661739)
  (3.76001,0.682115)(3.78001,0.705695)(3.80001,0.733341)
  (3.82001,0.766237)(3.84001,0.806043)(3.86001,0.855144)
  (3.88001,0.917039)(3.9,0.99696)(3.92,1.10275)(3.94,1.24578)
  (3.96,1.44028)(3.98,1.69613)(4,2.00004)(4.02,2.30509)(4.04,2.56337)
  (4.06,2.76022)(4.08,2.90499)(4.1,3.01195)(4.12,3.09262)
  (4.14,3.15498)(4.16,3.20434)(4.18,3.24427)(4.2,3.27718)
  (4.22,3.30475)(4.24,3.32819)(4.26,3.34836)(4.28,3.36592)
  (4.29999,3.38136)(4.31999,3.39504)(4.33999,3.40727)
  (4.35999,3.41828)(4.37999,3.42825)(4.39999,3.43733)
  (4.41999,3.44564)(4.43999,3.45329)(4.45999,3.46036)
  (4.47999,3.46693)(4.49999,3.47304)(4.51999,3.47876)
  (4.53999,3.48413)(4.55999,3.48918)(4.57999,3.49394)
  (4.59999,3.49846)(4.61999,3.50274)(4.63999,3.50681)(4.65999,3.5107)
  (4.67999,3.51442)(4.69998,3.51798)(4.71998,3.5214)(4.73998,3.52469)
  (4.75998,3.52786)(4.77998,3.53092)(4.79998,3.53388)
  (4.81998,3.53676)(4.83998,3.53954)(4.85998,3.54226)(4.87998,3.5449)
  (4.89998,3.54748)(4.91998,3.55)(4.93998,3.55246)(4.95998,3.55488)
  (4.97998,3.55725)(4.99998,3.55959)(5.01998,3.56189)
  (5.03998,3.56415)(5.05998,3.56639)(5.07998,3.5686)(5.09997,3.57079)
\path(5.10002,0.429206)(5.10902,0.430186)(5.11802,0.431163)
  (5.12702,0.432136)(5.13602,0.433107)(5.14502,0.434075)
  (5.15402,0.435041)(5.16302,0.436005)(5.17202,0.436967)
  (5.18102,0.437928)(5.19002,0.438887)(5.19902,0.439846)
  (5.20802,0.440804)(5.21702,0.441762)(5.22602,0.44272)
  (5.23502,0.443678)(5.24402,0.444636)(5.25302,0.445595)
  (5.26202,0.446556)(5.27102,0.447518)(5.28002,0.448481)
  (5.28902,0.449446)(5.29802,0.450413)(5.30702,0.451383)
  (5.31602,0.452356)(5.32502,0.453332)(5.33402,0.454311)
  (5.34302,0.455293)(5.35202,0.45628)(5.36102,0.457271)
  (5.37002,0.458267)(5.37902,0.459267)(5.38802,0.460273)
  (5.39702,0.461285)(5.40602,0.462302)(5.41502,0.463326)
  (5.42402,0.464356)(5.43302,0.465394)(5.44202,0.466439)
  (5.45102,0.467491)(5.46001,0.468552)(5.46901,0.469622)
  (5.47801,0.470701)(5.48701,0.471789)(5.49601,0.472887)
  (5.50501,0.473995)(5.51401,0.475115)(5.52301,0.476245)
  (5.53201,0.477388)(5.54101,0.478543)(5.55001,0.479711)
  (5.55901,0.480893)(5.56801,0.482088)(5.57701,0.483299)
  (5.58601,0.484525)(5.59501,0.485767)(5.60401,0.487025)
  (5.61301,0.488301)(5.62201,0.489595)(5.63101,0.490908)
  (5.64001,0.492241)(5.64901,0.493594)(5.65801,0.494969)
  (5.66701,0.496367)(5.67601,0.497787)(5.68501,0.499232)
  (5.69401,0.500702)(5.70301,0.502199)(5.71201,0.503724)
  (5.72101,0.505277)(5.73001,0.50686)(5.73901,0.508475)
  (5.74801,0.510123)(5.75701,0.511806)(5.76601,0.513524)
  (5.77501,0.51528)(5.78401,0.517075)(5.79301,0.518911)
  (5.80201,0.520791)(5.81101,0.522716)(5.82,0.524688)(5.829,0.52671)
  (5.838,0.528785)(5.847,0.530914)(5.856,0.533101)(5.865,0.535348)
  (5.874,0.537659)(5.883,0.540037)(5.892,0.542487)(5.901,0.54501)
  (5.91,0.547613)(5.919,0.550299)(5.928,0.553072)(5.937,0.555939)
  (5.946,0.558904)(5.955,0.561974)(5.964,0.565155)(5.973,0.568454)
  (5.982,0.571878)(5.991,0.575436)(6,0.579136)
\allinethickness{0.4pt}%
\color{rgb_000000}%
\path(0,3.5708)(0.0348837,3.5708)
\path(0.104651,3.5708)(0.139535,3.5708)
\path(0.139535,3.5708)(0.174419,3.5708)
\path(0.244186,3.5708)(0.27907,3.5708)
\path(0.27907,3.5708)(0.313953,3.5708)
\path(0.383721,3.5708)(0.418605,3.5708)
\path(0.418605,3.5708)(0.453488,3.5708)
\path(0.523256,3.5708)(0.55814,3.5708)
\path(0.55814,3.5708)(0.593023,3.5708)
\path(0.662791,3.5708)(0.697674,3.5708)
\path(0.697674,3.5708)(0.732558,3.5708)
\path(0.802326,3.5708)(0.837209,3.5708)
\path(0.837209,3.5708)(0.872093,3.5708)
\path(0.94186,3.5708)(0.976744,3.5708)
\path(0.976744,3.5708)(1.01163,3.5708)
\path(1.0814,3.5708)(1.11628,3.5708)
\path(1.11628,3.5708)(1.15116,3.5708)
\path(1.22093,3.5708)(1.25581,3.5708)
\path(1.25581,3.5708)(1.2907,3.5708)
\path(1.36047,3.5708)(1.39535,3.5708)
\path(1.39535,3.5708)(1.43023,3.5708)
\path(1.5,3.5708)(1.53488,3.5708)
\path(1.53488,3.5708)(1.56977,3.5708)
\path(1.63953,3.5708)(1.67442,3.5708)
\path(1.67442,3.5708)(1.7093,3.5708)
\path(1.77907,3.5708)(1.81395,3.5708)
\path(1.81395,3.5708)(1.84884,3.5708)
\path(1.9186,3.5708)(1.95349,3.5708)
\path(1.95349,3.5708)(1.98837,3.5708)
\path(2.05814,3.5708)(2.09302,3.5708)
\path(2.09302,3.5708)(2.12791,3.5708)
\path(2.19767,3.5708)(2.23256,3.5708)
\path(2.23256,3.5708)(2.26744,3.5708)
\path(2.33721,3.5708)(2.37209,3.5708)
\path(2.37209,3.5708)(2.40698,3.5708)
\path(2.47674,3.5708)(2.51163,3.5708)
\path(2.51163,3.5708)(2.54651,3.5708)
\path(2.61628,3.5708)(2.65116,3.5708)
\path(2.65116,3.5708)(2.68605,3.5708)
\path(2.75581,3.5708)(2.7907,3.5708)
\path(2.7907,3.5708)(2.82558,3.5708)
\path(2.89535,3.5708)(2.93023,3.5708)
\path(2.93023,3.5708)(2.96512,3.5708)
\path(3.03488,3.5708)(3.06977,3.5708)
\path(3.06977,3.5708)(3.10465,3.5708)
\path(3.17442,3.5708)(3.2093,3.5708)
\path(3.2093,3.5708)(3.24419,3.5708)
\path(3.31395,3.5708)(3.34884,3.5708)
\path(3.34884,3.5708)(3.38372,3.5708)
\path(3.45349,3.5708)(3.48837,3.5708)
\path(3.48837,3.5708)(3.52326,3.5708)
\path(3.59302,3.5708)(3.62791,3.5708)
\path(3.62791,3.5708)(3.66279,3.5708)
\path(3.73256,3.5708)(3.76744,3.5708)
\path(3.76744,3.5708)(3.80233,3.5708)
\path(3.87209,3.5708)(3.90698,3.5708)
\path(3.90698,3.5708)(3.94186,3.5708)
\path(4.01163,3.5708)(4.04651,3.5708)
\path(4.04651,3.5708)(4.0814,3.5708)
\path(4.15116,3.5708)(4.18605,3.5708)
\path(4.18605,3.5708)(4.22093,3.5708)
\path(4.2907,3.5708)(4.32558,3.5708)
\path(4.32558,3.5708)(4.36047,3.5708)
\path(4.43023,3.5708)(4.46512,3.5708)
\path(4.46512,3.5708)(4.5,3.5708)
\path(4.56977,3.5708)(4.60465,3.5708)
\path(4.60465,3.5708)(4.63953,3.5708)
\path(4.7093,3.5708)(4.74419,3.5708)
\path(4.74419,3.5708)(4.77907,3.5708)
\path(4.84884,3.5708)(4.88372,3.5708)
\path(4.88372,3.5708)(4.9186,3.5708)
\path(4.98837,3.5708)(5.02326,3.5708)
\path(5.02326,3.5708)(5.05814,3.5708)
\path(5.12791,3.5708)(5.16279,3.5708)
\path(5.16279,3.5708)(5.19767,3.5708)
\path(5.26744,3.5708)(5.30233,3.5708)
\path(5.30233,3.5708)(5.33721,3.5708)
\path(5.40698,3.5708)(5.44186,3.5708)
\path(5.44186,3.5708)(5.47674,3.5708)
\path(5.54651,3.5708)(5.5814,3.5708)
\path(5.5814,3.5708)(5.61628,3.5708)
\path(5.68605,3.5708)(5.72093,3.5708)
\path(5.72093,3.5708)(5.75581,3.5708)
\path(5.82558,3.5708)(5.86047,3.5708)
\path(5.86047,3.5708)(5.89535,3.5708)
\path(5.96512,3.5708)(6,3.5708)
\path(0,0.429204)(0.0348837,0.429204)
\path(0.104651,0.429204)(0.139535,0.429204)
\path(0.139535,0.429204)(0.174419,0.429204)
\path(0.244186,0.429204)(0.27907,0.429204)
\path(0.27907,0.429204)(0.313953,0.429204)
\path(0.383721,0.429204)(0.418605,0.429204)
\path(0.418605,0.429204)(0.453488,0.429204)
\path(0.523256,0.429204)(0.55814,0.429204)
\path(0.55814,0.429204)(0.593023,0.429204)
\path(0.662791,0.429204)(0.697674,0.429204)
\path(0.697674,0.429204)(0.732558,0.429204)
\path(0.802326,0.429204)(0.837209,0.429204)
\path(0.837209,0.429204)(0.872093,0.429204)
\path(0.94186,0.429204)(0.976744,0.429204)
\path(0.976744,0.429204)(1.01163,0.429204)
\path(1.0814,0.429204)(1.11628,0.429204)
\path(1.11628,0.429204)(1.15116,0.429204)
\path(1.22093,0.429204)(1.25581,0.429204)
\path(1.25581,0.429204)(1.2907,0.429204)
\path(1.36047,0.429204)(1.39535,0.429204)
\path(1.39535,0.429204)(1.43023,0.429204)
\path(1.5,0.429204)(1.53488,0.429204)
\path(1.53488,0.429204)(1.56977,0.429204)
\path(1.63953,0.429204)(1.67442,0.429204)
\path(1.67442,0.429204)(1.7093,0.429204)
\path(1.77907,0.429204)(1.81395,0.429204)
\path(1.81395,0.429204)(1.84884,0.429204)
\path(1.9186,0.429204)(1.95349,0.429204)
\path(1.95349,0.429204)(1.98837,0.429204)
\path(2.05814,0.429204)(2.09302,0.429204)
\path(2.09302,0.429204)(2.12791,0.429204)
\path(2.19767,0.429204)(2.23256,0.429204)
\path(2.23256,0.429204)(2.26744,0.429204)
\path(2.33721,0.429204)(2.37209,0.429204)
\path(2.37209,0.429204)(2.40698,0.429204)
\path(2.47674,0.429204)(2.51163,0.429204)
\path(2.51163,0.429204)(2.54651,0.429204)
\path(2.61628,0.429204)(2.65116,0.429204)
\path(2.65116,0.429204)(2.68605,0.429204)
\path(2.75581,0.429204)(2.7907,0.429204)
\path(2.7907,0.429204)(2.82558,0.429204)
\path(2.89535,0.429204)(2.93023,0.429204)
\path(2.93023,0.429204)(2.96512,0.429204)
\path(3.03488,0.429204)(3.06977,0.429204)
\path(3.06977,0.429204)(3.10465,0.429204)
\path(3.17442,0.429204)(3.2093,0.429204)
\path(3.2093,0.429204)(3.24419,0.429204)
\path(3.31395,0.429204)(3.34884,0.429204)
\path(3.34884,0.429204)(3.38372,0.429204)
\path(3.45349,0.429204)(3.48837,0.429204)
\path(3.48837,0.429204)(3.52326,0.429204)
\path(3.59302,0.429204)(3.62791,0.429204)
\path(3.62791,0.429204)(3.66279,0.429204)
\path(3.73256,0.429204)(3.76744,0.429204)
\path(3.76744,0.429204)(3.80233,0.429204)
\path(3.87209,0.429204)(3.90698,0.429204)
\path(3.90698,0.429204)(3.94186,0.429204)
\path(4.01163,0.429204)(4.04651,0.429204)
\path(4.04651,0.429204)(4.0814,0.429204)
\path(4.15116,0.429204)(4.18605,0.429204)
\path(4.18605,0.429204)(4.22093,0.429204)
\path(4.2907,0.429204)(4.32558,0.429204)
\path(4.32558,0.429204)(4.36047,0.429204)
\path(4.43023,0.429204)(4.46512,0.429204)
\path(4.46512,0.429204)(4.5,0.429204)
\path(4.56977,0.429204)(4.60465,0.429204)
\path(4.60465,0.429204)(4.63953,0.429204)
\path(4.7093,0.429204)(4.74419,0.429204)
\path(4.74419,0.429204)(4.77907,0.429204)
\path(4.84884,0.429204)(4.88372,0.429204)
\path(4.88372,0.429204)(4.9186,0.429204)
\path(4.98837,0.429204)(5.02326,0.429204)
\path(5.02326,0.429204)(5.05814,0.429204)
\path(5.12791,0.429204)(5.16279,0.429204)
\path(5.16279,0.429204)(5.19767,0.429204)
\path(5.26744,0.429204)(5.30233,0.429204)
\path(5.30233,0.429204)(5.33721,0.429204)
\path(5.40698,0.429204)(5.44186,0.429204)
\path(5.44186,0.429204)(5.47674,0.429204)
\path(5.54651,0.429204)(5.5814,0.429204)
\path(5.5814,0.429204)(5.61628,0.429204)
\path(5.68605,0.429204)(5.72093,0.429204)
\path(5.72093,0.429204)(5.75581,0.429204)
\path(5.82558,0.429204)(5.86047,0.429204)
\path(5.86047,0.429204)(5.89535,0.429204)
\path(5.96512,0.429204)(6,0.429204)
\path(1.9,0)(1.9,0.0344828)
\path(1.9,0.103448)(1.9,0.137931)
\path(1.9,0.137931)(1.9,0.172414)
\path(1.9,0.241379)(1.9,0.275862)
\path(1.9,0.275862)(1.9,0.310345)
\path(1.9,0.37931)(1.9,0.413793)
\path(1.9,0.413793)(1.9,0.448276)
\path(1.9,0.517241)(1.9,0.551724)
\path(1.9,0.551724)(1.9,0.586207)
\path(1.9,0.655172)(1.9,0.689655)
\path(1.9,0.689655)(1.9,0.724138)
\path(1.9,0.793103)(1.9,0.827586)
\path(1.9,0.827586)(1.9,0.862069)
\path(1.9,0.931034)(1.9,0.965517)
\path(1.9,0.965517)(1.9,1)
\path(1.9,1.06897)(1.9,1.10345)
\path(1.9,1.10345)(1.9,1.13793)
\path(1.9,1.2069)(1.9,1.24138)
\path(1.9,1.24138)(1.9,1.27586)
\path(1.9,1.34483)(1.9,1.37931)
\path(1.9,1.37931)(1.9,1.41379)
\path(1.9,1.48276)(1.9,1.51724)
\path(1.9,1.51724)(1.9,1.55172)
\path(1.9,1.62069)(1.9,1.65517)
\path(1.9,1.65517)(1.9,1.68966)
\path(1.9,1.75862)(1.9,1.7931)
\path(1.9,1.7931)(1.9,1.82759)
\path(1.9,1.89655)(1.9,1.93103)
\path(1.9,1.93103)(1.9,1.96552)
\path(1.9,2.03448)(1.9,2.06897)
\path(1.9,2.06897)(1.9,2.10345)
\path(1.9,2.17241)(1.9,2.2069)
\path(1.9,2.2069)(1.9,2.24138)
\path(1.9,2.31034)(1.9,2.34483)
\path(1.9,2.34483)(1.9,2.37931)
\path(1.9,2.44828)(1.9,2.48276)
\path(1.9,2.48276)(1.9,2.51724)
\path(1.9,2.58621)(1.9,2.62069)
\path(1.9,2.62069)(1.9,2.65517)
\path(1.9,2.72414)(1.9,2.75862)
\path(1.9,2.75862)(1.9,2.7931)
\path(1.9,2.86207)(1.9,2.89655)
\path(1.9,2.89655)(1.9,2.93103)
\path(1.9,3)(1.9,3.03448)
\path(1.9,3.03448)(1.9,3.06897)
\path(1.9,3.13793)(1.9,3.17241)
\path(1.9,3.17241)(1.9,3.2069)
\path(1.9,3.27586)(1.9,3.31034)
\path(1.9,3.31034)(1.9,3.34483)
\path(1.9,3.41379)(1.9,3.44828)
\path(1.9,3.44828)(1.9,3.48276)
\path(1.9,3.55172)(1.9,3.58621)
\path(1.9,3.58621)(1.9,3.62069)
\path(1.9,3.68966)(1.9,3.72414)
\path(1.9,3.72414)(1.9,3.75862)
\path(1.9,3.82759)(1.9,3.86207)
\path(1.9,3.86207)(1.9,3.89655)
\path(1.9,3.96552)(1.9,4)
\path(3.1,0)(3.1,0.0344828)
\path(3.1,0.103448)(3.1,0.137931)
\path(3.1,0.137931)(3.1,0.172414)
\path(3.1,0.241379)(3.1,0.275862)
\path(3.1,0.275862)(3.1,0.310345)
\path(3.1,0.37931)(3.1,0.413793)
\path(3.1,0.413793)(3.1,0.448276)
\path(3.1,0.517241)(3.1,0.551724)
\path(3.1,0.551724)(3.1,0.586207)
\path(3.1,0.655172)(3.1,0.689655)
\path(3.1,0.689655)(3.1,0.724138)
\path(3.1,0.793103)(3.1,0.827586)
\path(3.1,0.827586)(3.1,0.862069)
\path(3.1,0.931034)(3.1,0.965517)
\path(3.1,0.965517)(3.1,1)
\path(3.1,1.06897)(3.1,1.10345)
\path(3.1,1.10345)(3.1,1.13793)
\path(3.1,1.2069)(3.1,1.24138)
\path(3.1,1.24138)(3.1,1.27586)
\path(3.1,1.34483)(3.1,1.37931)
\path(3.1,1.37931)(3.1,1.41379)
\path(3.1,1.48276)(3.1,1.51724)
\path(3.1,1.51724)(3.1,1.55172)
\path(3.1,1.62069)(3.1,1.65517)
\path(3.1,1.65517)(3.1,1.68966)
\path(3.1,1.75862)(3.1,1.7931)
\path(3.1,1.7931)(3.1,1.82759)
\path(3.1,1.89655)(3.1,1.93103)
\path(3.1,1.93103)(3.1,1.96552)
\path(3.1,2.03448)(3.1,2.06897)
\path(3.1,2.06897)(3.1,2.10345)
\path(3.1,2.17241)(3.1,2.2069)
\path(3.1,2.2069)(3.1,2.24138)
\path(3.1,2.31034)(3.1,2.34483)
\path(3.1,2.34483)(3.1,2.37931)
\path(3.1,2.44828)(3.1,2.48276)
\path(3.1,2.48276)(3.1,2.51724)
\path(3.1,2.58621)(3.1,2.62069)
\path(3.1,2.62069)(3.1,2.65517)
\path(3.1,2.72414)(3.1,2.75862)
\path(3.1,2.75862)(3.1,2.7931)
\path(3.1,2.86207)(3.1,2.89655)
\path(3.1,2.89655)(3.1,2.93103)
\path(3.1,3)(3.1,3.03448)
\path(3.1,3.03448)(3.1,3.06897)
\path(3.1,3.13793)(3.1,3.17241)
\path(3.1,3.17241)(3.1,3.2069)
\path(3.1,3.27586)(3.1,3.31034)
\path(3.1,3.31034)(3.1,3.34483)
\path(3.1,3.41379)(3.1,3.44828)
\path(3.1,3.44828)(3.1,3.48276)
\path(3.1,3.55172)(3.1,3.58621)
\path(3.1,3.58621)(3.1,3.62069)
\path(3.1,3.68966)(3.1,3.72414)
\path(3.1,3.72414)(3.1,3.75862)
\path(3.1,3.82759)(3.1,3.86207)
\path(3.1,3.86207)(3.1,3.89655)
\path(3.1,3.96552)(3.1,4)
\path(5.1,0)(5.1,0.0344828)
\path(5.1,0.103448)(5.1,0.137931)
\path(5.1,0.137931)(5.1,0.172414)
\path(5.1,0.241379)(5.1,0.275862)
\path(5.1,0.275862)(5.1,0.310345)
\path(5.1,0.37931)(5.1,0.413793)
\path(5.1,0.413793)(5.1,0.448276)
\path(5.1,0.517241)(5.1,0.551724)
\path(5.1,0.551724)(5.1,0.586207)
\path(5.1,0.655172)(5.1,0.689655)
\path(5.1,0.689655)(5.1,0.724138)
\path(5.1,0.793103)(5.1,0.827586)
\path(5.1,0.827586)(5.1,0.862069)
\path(5.1,0.931034)(5.1,0.965517)
\path(5.1,0.965517)(5.1,1)
\path(5.1,1.06897)(5.1,1.10345)
\path(5.1,1.10345)(5.1,1.13793)
\path(5.1,1.2069)(5.1,1.24138)
\path(5.1,1.24138)(5.1,1.27586)
\path(5.1,1.34483)(5.1,1.37931)
\path(5.1,1.37931)(5.1,1.41379)
\path(5.1,1.48276)(5.1,1.51724)
\path(5.1,1.51724)(5.1,1.55172)
\path(5.1,1.62069)(5.1,1.65517)
\path(5.1,1.65517)(5.1,1.68966)
\path(5.1,1.75862)(5.1,1.7931)
\path(5.1,1.7931)(5.1,1.82759)
\path(5.1,1.89655)(5.1,1.93103)
\path(5.1,1.93103)(5.1,1.96552)
\path(5.1,2.03448)(5.1,2.06897)
\path(5.1,2.06897)(5.1,2.10345)
\path(5.1,2.17241)(5.1,2.2069)
\path(5.1,2.2069)(5.1,2.24138)
\path(5.1,2.31034)(5.1,2.34483)
\path(5.1,2.34483)(5.1,2.37931)
\path(5.1,2.44828)(5.1,2.48276)
\path(5.1,2.48276)(5.1,2.51724)
\path(5.1,2.58621)(5.1,2.62069)
\path(5.1,2.62069)(5.1,2.65517)
\path(5.1,2.72414)(5.1,2.75862)
\path(5.1,2.75862)(5.1,2.7931)
\path(5.1,2.86207)(5.1,2.89655)
\path(5.1,2.89655)(5.1,2.93103)
\path(5.1,3)(5.1,3.03448)
\path(5.1,3.03448)(5.1,3.06897)
\path(5.1,3.13793)(5.1,3.17241)
\path(5.1,3.17241)(5.1,3.2069)
\path(5.1,3.27586)(5.1,3.31034)
\path(5.1,3.31034)(5.1,3.34483)
\path(5.1,3.41379)(5.1,3.44828)
\path(5.1,3.44828)(5.1,3.48276)
\path(5.1,3.55172)(5.1,3.58621)
\path(5.1,3.58621)(5.1,3.62069)
\path(5.1,3.68966)(5.1,3.72414)
\path(5.1,3.72414)(5.1,3.75862)
\path(5.1,3.82759)(5.1,3.86207)
\path(5.1,3.86207)(5.1,3.89655)
\path(5.1,3.96552)(5.1,4)
\put(1.5,1.09498){\color{rgb_ff0000}$\allinethickness{0.052719cm}\circle{0.052719}$}%
\put(1.46485,2.03515){\makebox(0,0)[br]{\hbox{\color{rgb_000000}\scriptsize $0$}}}
\put(1.5,4.07029){\makebox(0,0)[b]{\hbox{\color{rgb_000000}\scriptsize $\alpha_{\pm}$}}}
\put(6.07029,2){\makebox(0,0)[l]{\hbox{\color{rgb_000000}\scriptsize $\lambda$}}}
\put(1.42971,1.09498){\makebox(0,0)[r]{\hbox{\color{rgb_000000}\scriptsize \color{red}{$\alpha_{\ast}$}}}}
\put(1.93515,2.03565){\makebox(0,0)[bl]{\hbox{\color{rgb_000000}\scriptsize $4^{2}$}}}
\put(3.13515,2.03565){\makebox(0,0)[bl]{\hbox{\color{rgb_000000}\scriptsize $8^{2}$}}}
\put(5.13515,2.03565){\makebox(0,0)[bl]{\hbox{\color{rgb_000000}\scriptsize $12^{2}$}}}
\put(1.93512,3.60592){\makebox(0,0)[bl]{\hbox{\color{rgb_000000}\scriptsize \textcolor{blue}{$\lambda_{\pm,0}$}}}}
\put(3.13512,3.60594){\makebox(0,0)[bl]{\hbox{\color{rgb_000000}\scriptsize \textcolor{blue}{$\lambda_{\pm,1}$}}}}
\put(5.13512,3.60594){\makebox(0,0)[bl]{\hbox{\color{rgb_000000}\scriptsize \textcolor{blue}{$\lambda_{\pm,2}$}}}}
\put(1.46485,0.394058){\makebox(0,0)[tr]{\hbox{\color{rgb_000000}\scriptsize $-\pi$}}}
\put(1.46485,3.60594){\makebox(0,0)[br]{\hbox{\color{rgb_000000}\scriptsize $+\pi$}}}
\end{picture}%